\documentclass[prl,aps,reprint,twocolumn,footinbib,showpacs,superscriptaddress]{revtex4-1}
\usepackage{graphicx}
\usepackage{amssymb}
\usepackage{amsmath}
\usepackage{times}
\usepackage{bm}

\usepackage{color} %

\begin{document}

\title{Diverging fluctuations of the Lyapunov exponents}

 \author{Diego Paz\'o}
 \email{pazo@ifca.unican.es}
 \affiliation{Instituto de F\'{\i}sica de Cantabria (IFCA), CSIC-Universidad de
 Cantabria, 39005 Santander, Spain}
 \author{Juan M. L\'opez}
 \email{lopez@ifca.unican.es}
 \affiliation{Instituto de F\'{\i}sica de Cantabria (IFCA), CSIC-Universidad de
 Cantabria, 39005 Santander, Spain}
 \author{Antonio Politi}
\email{a.politi@abdn.ac.uk}
\affiliation{Institute for Complex Systems and Mathematical Biology and SUPA,
University of Aberdeen, Aberdeen AB24 3UE, United Kingdom}

\date{\today}

\begin{abstract}
We show that in generic one-dimensional Hamiltonian lattices the diffusion
coefficient of the maximum Lyapunov exponent diverges in the thermodynamic
limit. We trace this back to the long-range correlations associated with the
evolution of the hydrodynamic modes. In the case of normal heat transport, the
divergence is even stronger, leading to the breakdown of the usual
single-function Family-Vicsek scaling ansatz. A similar scenario is expected to
arise in the evolution of rough interfaces in the presence of a suitably
correlated background noise.
\end{abstract}

\maketitle

Lyapunov exponents (LEs) are dynamical invariants that provide a detailed
characterization of low-dimensional as well as spatio-temporal
chaos~\cite{Pikovsky}: they indeed allow estimating the fractal dimension, the
Kolmogorov-Sinai entropy and to ascertain extensivity of the
underlying dynamical regime. LEs are average quantities, defined as the
infinite-time limit of the so-called finite-time Lyapunov exponents (FTLEs).
Interestingly, also the temporal fluctuations of FTLEs carry important
information that is ultimately encoded in yet another
invariant: a suitable large deviation function. Fluctuations help to shed light
on important phenomena such as intermittency, strange nonchaotic attractors, and
stable chaos~\cite{Pikovsky}.

In dissipative systems with many degrees of freedom, the fluctuations of the
largest FTLE have been investigated in various numerical setups such as a shell
model for the energy cascade in turbulence~\cite{crisanti93}, a cellular
automaton~\cite{crisanti92}, molecular dynamics simulations~\cite{green13},
coupled-map lattice models~\cite{kuptsov11,laffargue13}, and a variety of
continuous-time models~\cite{nakao03,plp13}. In particular, in spatially
extended systems like those in \cite{kuptsov11,laffargue13,plp13}, the dynamics 
of Lyapunov vectors, i.e. perturbation fields, is
formally equivalent to the evolution of rough interfaces in a noisy
environment, the LE corresponding to the velocity of the interface
\cite{pik94,pik98}. This relationship is essentially based on the interpretation
of the logarithm of the local amplitude of the perturbation with the height
$h(x,t)$ of a suitable interface. As a result, the same ``physics'' can be found
in two significantly different contexts. In particular, the universality class
of roughening phenomena identified by the Kardar-Parisi-Zhang (KPZ) equation
\cite{kpz} includes also the perturbation evolution in spatially extended
chaotic systems \cite{pik94,pik98,pazo_lopez10}.

In spite of its broadness, the KPZ universality class does not encompass
Hamiltonian models~\cite{pik01,romero10}. Preliminary studies revealed different
critical properties and attributed the anomalous scaling to non-better specified
long-range correlations~\cite{pik01}. Later on, powerful methods for the
characterization of large deviations revealed that extreme fluctuations of the
FTLE in the classical Fermi-Pasta-Ulam (FPU-$\beta$) chain~\cite{FPU} correspond
to atypical solutions of soliton-like and chaotic-breather
dynamics~\cite{tailleur07,laffargue13}. However, it is not clear to what extent
they are responsible for the anomalous non-KPZ behavior.

In this Letter we study the diffusion coefficient $D$ of the maximum FTLE
in two prototypical Hamiltonian lattices (the FPU-$\beta$ and the $\Phi^4$ models).
Contrary to what observed in dissipative dynamics, where $D$ vanishes in
the thermodynamic limit, here $D$ diverges.
Within the rough-interface context, this behavior means that the velocity
of the interface does not self-average.
Otherwise said, the infinite-time limit (implied by the definition of the LE)
does not commute with the thermodynamic limit. This is yet another example of
how subtle the interrelations between these two limits may be 
(see also the Hamiltonian mean-field model, where an exchange of limits even 
transforms a vanishing into a finite
LE~\cite{Ginelli-Takeuchi-11}; or stable
chaos~\cite{Politi-Torcini-10}, where the non-commutation of the two limits is
at the origin of a self-sustained irregular dynamics in linearly stable
environments). Here we show that the divergence of the fluctuations originates
from the long-range spatio-temporal correlations that are naturally present
in Hamiltonian systems because of conservation laws (notably energy conservation).
A similar scenario is expected to arise in the evolution of rough interfaces in 
the presence of a suitably correlated background noise. 

Our results complement the pioneering work by McNamara and
Mareschal \cite{Mcnamara-01}, who established a connection between hydrodynamics 
and Lyapunov dynamics, by analyzing the evolution of the Lyapunov vectors
associated to small LEs. Here we shed further light, showing that
hydrodynamics shapes the most unstable direction as well.

\paragraph{Theory.--} Given an infinitesimal perturbation ${\bm{\delta u}}(t)$
pointing along the most unstable direction in tangent space (the so-called
leading Lyapunov vector), we denote its expansion factor over a time $t$ by
$\mathrm{e}^{\Gamma(t)}$. The ratio $\lambda(t) = \Gamma(t)/t$, the so-called
FTLE, is expected to fluctuate because of the heterogeneity of the degree of instability
across phase space. The minimal way to gauge the fluctuations
of the FTLE is through the variance
\begin{equation}
\chi^2(t) =\left< \left( \Gamma{\color{black}(t)} - \langle\lambda\rangle t \right)^2 \right> \; ,
\label{chi2}
\end{equation}
where the angular bracket $\langle\cdot\rangle$ denotes an average over an
ensemble of trajectories, and $\langle\lambda\rangle$ 
{\color{black} coincides with} the LE 
{\color{black} $\lim_{t\to\infty} \lambda(t)$
under the assumption that there is only one ergodic component for the energies 
considered.}

The fluctuations of the FTLE are quantified by the diffusion coefficient
\begin{equation}
D=\lim_{t\to\infty} \frac{\chi^2}{t},
\end{equation}
which is itself a dynamical invariant {\color{black} (i.e. independent of the norm 
type used to compute $\lambda(t)$)}.
In extended dynamical systems the diffusion coefficient is expected to
scale with the system size $L$ as
$D\sim L^{-\gamma}$ \cite{kuptsov11,plp13}, where $\gamma$ is called wandering exponent.

The scaling behavior of $D$ can be better understood by interpreting the
logarithm of the local amplitude of the perturbation $|\delta u_i|$ as the
height of a (rough) surface~\cite{pik94,pik98}: $h_i = \log |\delta u_i|$. Once
introduced the auxiliary field  $\phi_i(t) = h_i(t)-h_i(0)$, its spatial average
$\bar\phi(t)=(1/L)\sum_{i=1}^{L}\phi_i(t)$ corresponds to
$\Gamma(t)$~\footnote{Once the geometric norm $\lVert{\bm{\delta u}}\rVert_0 =
\prod_{i=1}^{L} \left|\delta u_i\right|^{1/L}$ is adopted.}, so that the FTLE
coincides with average velocity of the interface.

In the theory of roughening processes one observable of great interest is the
(squared) width of the interface $W^2(t) = \left<
\overline{\left(\phi{\color{black}_i}-\bar\phi\right)^2}\right>$, which, for self-affine
interfaces, satisfies the Family-Vicsek scaling ansatz
\begin{equation}
W^2 = L^{2\alpha} {\cal F}(t/L^z) \; ,
\label{ansatz_w2}
\end{equation}
where $\alpha$ and $z$ are the usual roughness and dynamical exponents,
respectively, and ${\cal F}(u)=\mathrm{const.}$ for $u\gg1$. The validity of
this relationship in the context of Lyapunov dynamics in extended dissipative
dynamical has been repeatedly
investigated~\cite{pik94,pik98,szendro07,pazo08,pazo_lopez10}, showing that the
leading Lyapunov vector falls within the universality class of KPZ
dynamics~\cite{kpz}.

The observable $\chi^2$ defined in Eq.~(\ref{chi2}) is, within the surface
framework, given by 
$\chi^2(t) =\left< \left( \bar\phi - \left<\bar\phi\right> \right)^2 \right>$.
For $\chi^2$ it is legitimate to invoke again a scaling ansatz \cite{plp13}
\begin{equation}
\chi^2 = L^{2\alpha} {\cal G}(t/L^z) (t/L^z) \; ,
\label{ansatz_chi}
\end{equation}
where the explicit time dependence has been included to stress the asymptotic
linear growth of $\chi^2$ (${\cal G}(u)=\mathrm{const.}$ for $u\gg1$).
As a result, the wandering exponent is~\cite{plp13}
\begin{equation}
\gamma= z - 2\alpha  \; ,
\label{main}
\end{equation}
for any spatial dimension. In the case of chaotic dissipative systems, $\gamma$
is universal. In fact, as the relationship with KPZ dynamics holds $\alpha=
1/2$, $z=3/2$ in one dimension, so that $\gamma=1/2$; analogously, $\gamma
\approx 0.839$ in two dimensions~\cite{plp13}. In both cases $\gamma > 0$
implies that the fluctuations of the FTLE decrease upon increasing the system
size, thereby indicating that the LE self-averages in the thermodynamic limit. 

\paragraph{Models.--}
The only set of systems with spatio-temporal (extensive) chaos where the correspondence with KPZ
does not apply is the important class of Hamiltonian models~\cite{pik01}.
We investigate two popular Hamiltonian lattices: (i) the FPU-$\beta$ model, defined by the
evolution equation $\ddot q_i =  F(q_{i+1}-q_i) - F(q_{i}-q_{i-1})$
where $F(x) = x + x^3$, and by the tangent
space dynamics $\ddot Q_i = m_{i+1}(Q_{i+1}-Q_i) + m_i (Q_{i-1}-Q_i)$, where
$Q_i$ is the infinitesimal variation of $q_i$ and
$m_i = 1+ 3(q_i-q_{i-1})^2$ is the local multiplier;
(ii) the $\Phi^4$ model, in which case
$\ddot q_i =  q_{i+1}-q_i+q_{i-1}- q_i^3$, and
$\ddot Q_i = Q_{i+1}-m_iQ_i+Q_{i-1}$, where $m_i= 1+3q_i^2$.
In both cases the interface height is defined as
$h_i(t)=\ln|(Q_i^2(t)+\dot Q_i^2(t))|$. 
Periodic boundary conditions are always assumed and the equations are integrated
by using the McLachlan-Atela algorithm~\cite{ma}.
All simulations were carried out with the moderately high energy density
$E/L=5${\color{black}, which is above the strong stochasticity threshold.}

\begin{figure}
\centerline{\includegraphics[width=80mm,clip]{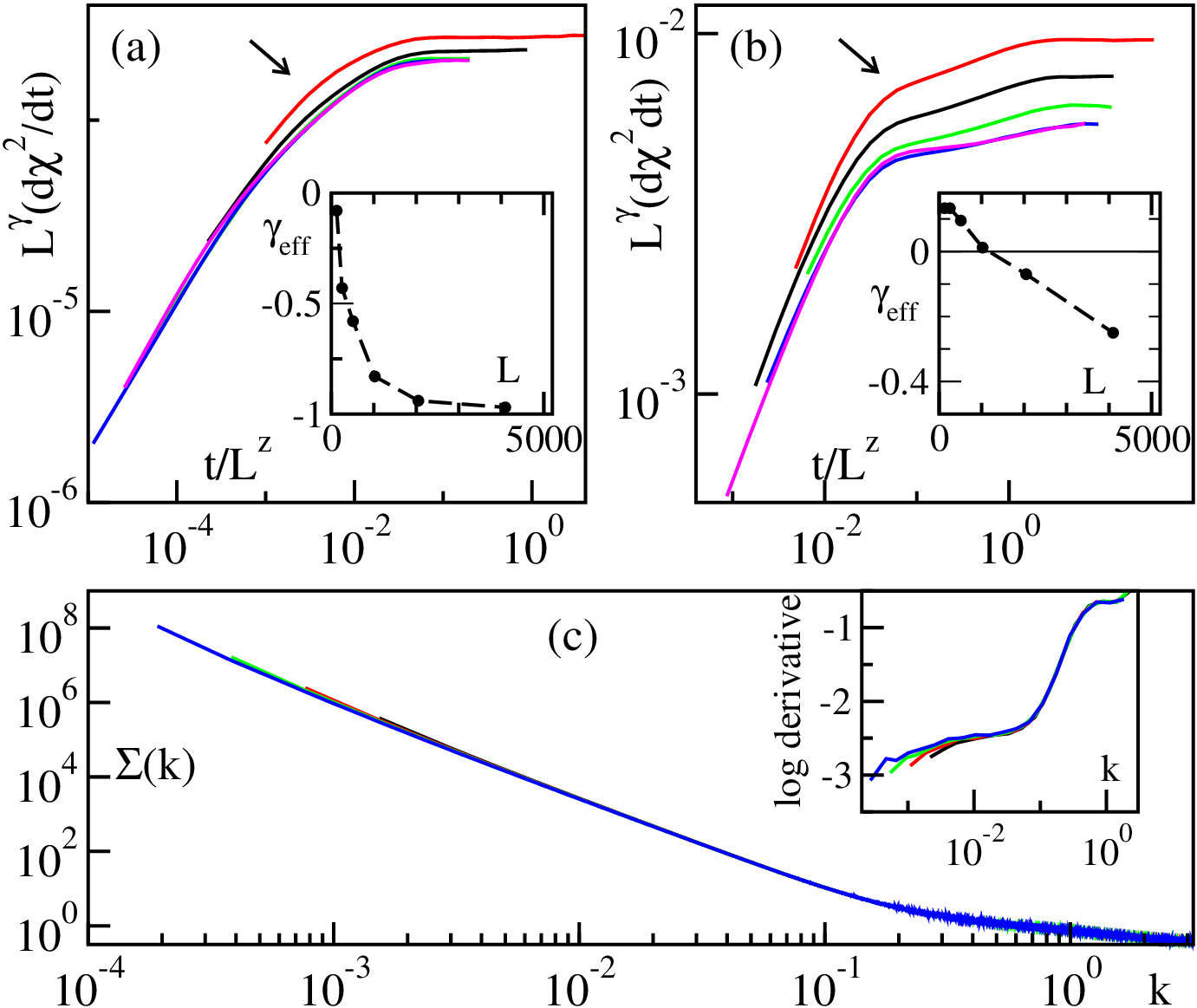}}
\caption{Rescaled FTLE fluctuations for the $\Phi^4$ (panel a) and FPU (panel b)
models, according to the ansatz \eqref{ansatz_chi}, for sizes $L=256,512,1024,2048,4096$.
The derivative $d\chi^2/dt$ is considered because of the faster temporal convergence to
the asymptotic results.
The optimal collapse of the data sets for the larger system sizes is achieved setting
$\gamma=-0.97$ and $z=2.15$ in (a) and $\gamma=-0.25$ and $z=1.45$ in (b).
The insets show the convergence of $\gamma$, determined by comparing the sizes $L$ and $L/2$.
In panel (c) the structure factor of Lyapunov-vector surface the FPU model
is plotted for $L=4096$, 8192, 16384, and 32768 (see black, red, green, and blue lines).
The logarithmic derivative is plotted in the inset, adopting the same color coding.}
\label{fig:fig1}
\end{figure}

The scaling of the diffusion coefficient $D$ with $L$ has been determined by
integrating the equations in tangent space and measuring $\chi^2$, as defined by
Eq.~\eqref{chi2}, for different system sizes. The results for the $\Phi^4$ model are shown
in Fig.~\ref{fig:fig1}(a) after a proper rescaling to conform to the scaling
ansatz~\eqref{ansatz_chi}. There we see that the agreement increases upon
increasing the system size with a clear evidence that $z\approx 2$. As for
$\gamma$, it is convincingly negative, but an accurate estimate of the
asymptotic value is problematic due to the slow convergence with the system
size. The effective wandering exponent $\gamma_{\mathrm{eff}}(L)$, obtained by
comparing the data for $L$ with that for $L/2$, is plotted in the inset in
Fig.~\ref{fig:fig1}(a). As a result, we can argue that $\gamma \to \approx -1$.

The results for the FPU-$\beta$ model are plotted in Fig.~\ref{fig:fig1}(b). On
the one hand, a robust estimate of $z \approx 3/2$ is found for the whole set of
system sizes considered in our simulations. On the other hand $\gamma$ varies
with $L$ and makes the estimate of the asymptotic value even more troublesome
than in the previous case. The analysis of the effective wandering exponent
$\gamma_{\mathrm{eff}}(L)$, plotted in the inset in Fig.~\ref{fig:fig1}(b),
suggests that $\gamma$ is at least more negative than $-0.25$. Given the strong
finite-size corrections, we have estimated $\alpha$ independently, from the
scaling behavior of the structure factor $\Sigma(k)=\langle|\hat
h_k(t)|^2\rangle$, where $\hat h_k(t)$ is the Fourier transform of the interface
profile at time $t$. The structure factor for $L=32768$, a fairly large lattice,
is shown in Fig.~\ref{fig:fig1}(c) where it appears power-law-like. By virtue of
Parseval's theorem one expects $\Sigma(k)\sim k^{-(2\alpha+1)}$. The inset
suggests that there is a very slow convergence to $\alpha=1$ in the
thermodynamic limit. This leads us to conjecture the set of exponents for the
FPU-$\beta$ model: $\alpha=1$, $z=3/2$, and hence, from Eq.~\eqref{main}, the
wandering exponent $\gamma = -1/2$. In sum, the relative fluctuations of the
FTLE diverge with $L$ in both models, though with different $\gamma$ values.

\begin{figure}
\centerline{\includegraphics[width=80mm]{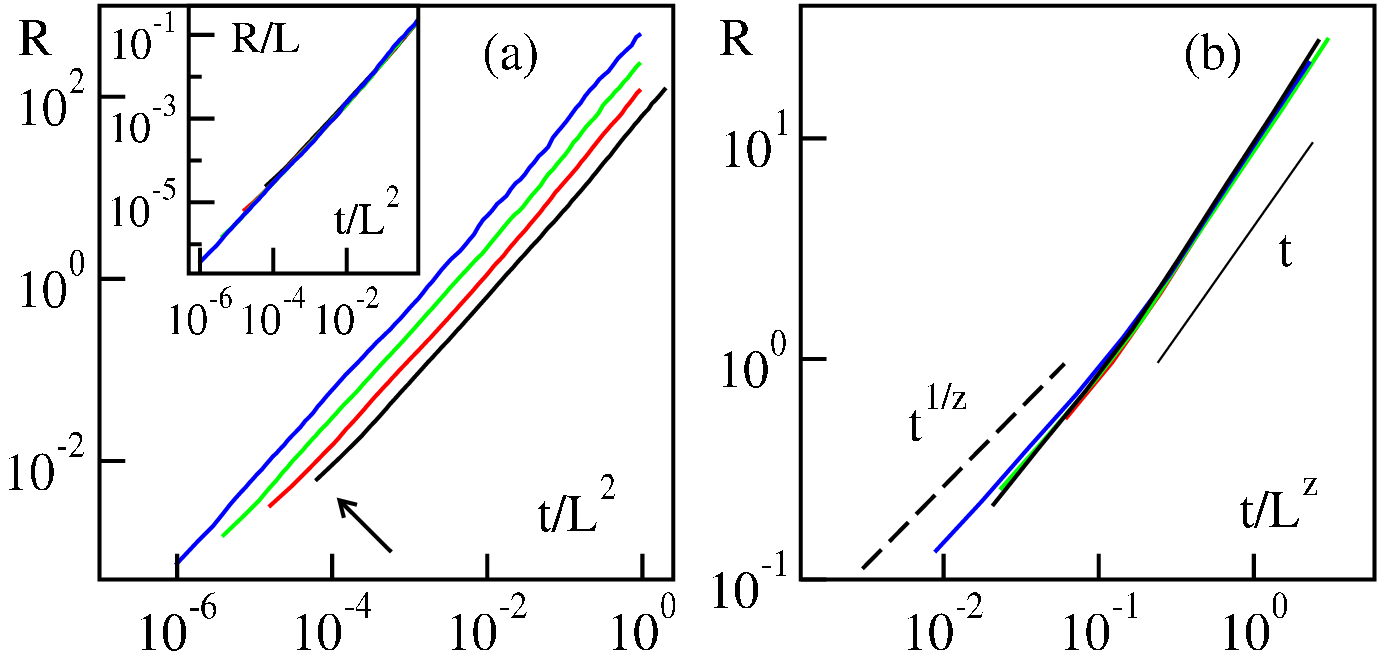}}
\caption{Evolution of $R$ (see Eq.~\eqref{R}) in the
$\Phi^4$ (panel (a)) and FPU (panel (b)) models, for
$L=512$, 1024, 2048, and 4096. $R/L$ is plotted in the inset of panel 
(a). A good data collapse is achieved for $z=1.4$ for the
FPU model.}
\label{fig:fig2}
\end{figure}

Given the peculiarities found, we decided to deepen the numerical analysis
by looking at the overall issue in a different way. More precisely, we have monitored
the ratio between the fluctuations of the FTLE and those of the interface width,
\begin{equation}
R = \chi^2/W^2 \; .
\label{R}
\end{equation}
>From Eqs.~(\ref{ansatz_w2}) and (\ref{ansatz_chi}), this dimensionless
observable is expected to be independent of $L$ if plotted versus the rescaled
time $u = t/L^z$. In Fig.~\ref{fig:fig2}(b) we see that this is indeed the case
for the FPU-$\beta$ model (there we have assumed $z=1.4$, not far from the value
$z=1.45$ estimated from Fig.~\ref{fig:fig1}(b). There, we also see that $R$
diverges linearly for large $t$, while it grows as $R \sim t^{1/z}$ at short
times, consistently with the scaling Ans\"atze in Eqs.~(\ref{ansatz_w2}) and
(\ref{ansatz_chi}) \footnote{For the norm type considered (and defined by the
field $\phi$) ${\cal G}(u\ll1)=u^{(2\alpha+1)/z}$~\cite{plp13}, while ${\cal
F}(u\ll1)=u^{2\alpha/z}$~\cite{Barabasi}.}, see Ref.~\cite{plp13}.

A completely different scenario is instead found for the $\Phi^4$ model. The
various curves reported in Fig.~\ref{fig:fig2}(a) do not collapse onto one
another (neither at short nor at long times). Additionally the time dependence
is linear all along the entire range. A nice data collapse is obtained only
after rescaling $R$ by $L$ (see the inset). The only way we have found to
reconcile this result with the two initial scaling hypotheses is by assuming the
existence, for the $\Phi^4$ system, of two different $\alpha$-exponents,
$\alpha_\chi$ and $\alpha_W$, in Eqs.~\eqref{ansatz_chi} and \eqref{ansatz_w2},
respectively, so that $R\sim L^{2(\alpha_\chi-\alpha_W)}(t/L^z)$. The observed
data collapse for the $R$-curves implies $\alpha_\chi = \alpha_W +1/2$.  An
independent study of the scaling of $W^2$ points to $\alpha_W=1$ in the
thermodynamic limit, whereupon $\alpha_\chi=3/2$. As a result, the relationship
\eqref{main} is still valid, once the proper $\alpha$ is being invoked:
$\gamma=z-2\alpha_\chi$. This yields $\gamma=-1$, in agreement with the direct
simulations in Fig.~\ref{fig:fig1}(a).

\paragraph{Stochastic model.--}
The study of the two Hamiltonian models has revealed a diverging diffusion
coefficient of the FTLE fluctuations as well as two significantly different
scaling scenarios. The recent progress in the thermodynamic properties of
oscillator chains has shown that the two models belong to different universality
classes: Fourier law is satisfied in the $\Phi^4$ model, while a divergence of
transport coefficients is found in the FPU-$\beta$ system~\cite{Lepri16}. More
precisely, the hydrodynamic behavior of $\Phi^4$ is a pure diffusion and
therefore characterized by $z=2$; while the scenario is more complex in the case
of FPU-type models, where $z$ depends on the symmetry of the interactions (see
\cite{Spohn2016}). Our results show that such a difference manifests
itself also in the context of Lyapunov dynamics.

In order to test to what extent the tangent-space dynamics is determined
by the correlation properties of the local multipliers $m_i(t)$,
we have studied the simple model
\begin{equation}
\delta u_i(t+1) = m_i(t)\left[\delta u_{i-1}(t)+\delta u_i(t)+\delta u_{i+1}(t)\right]\, ,
\label{stoch}
\end{equation}
where time is discrete and $m_i(t)$ is a stochastic term. It corresponds to the 
tangent space evolution of a generic coupled-map lattice.  If $m_i(t)$ is 
$\delta$-correlated both in space and time, the dynamics of $h_i=\log |\delta 
u_i|$ belongs to the KPZ universality class~\cite{pik94}. {\color{black} In 
contrast, for the $\Phi^4$ model we find that the  spectral density of the 
multipliers is
\begin{equation}
 \left<|\hat m_{k}(\omega)|^2 \right>= \frac{Ak^2}{Bk^4 
+ \omega^2} \; ,
\label{eq:sf_stoch}
\end{equation}
up to some finite-size corrections (data not shown).
The form of the hydrodynamic fluctuations given by
Eq.~\eqref{eq:sf_stoch} corresponds to diffusive 
transport~\cite{Forster}. Indeed, 
energy fluctuations relax diffusively in the $\Phi^4$ model~\cite{Lepri16} and
this is also expected for other observables like the local 
multipliers $m_i(t)$.}
In Ref.~\cite{Kipnis} it was  proposed a simple recipe to generate a
stochastic process $m_i$ characterized by the
{\color{black} spectral density}
in Eq.~(\ref{eq:sf_stoch}). Given a positive-defined field $m_i(t)$, a pair of
neighboring sites $i,i+1$ is randomly selected and the 
{\color{black} conserved quantity} $E_i=
m_i(t)+m_{i+1}(t)$ randomly redistributed over the two sites with a uniform
probability density in $[0,E_i]$ (so that detailed balance is satisfied). A time
unit corresponds to the performance of $L$ random moves. 

\begin{figure}
\centerline{\includegraphics[width=80mm,clip=true]{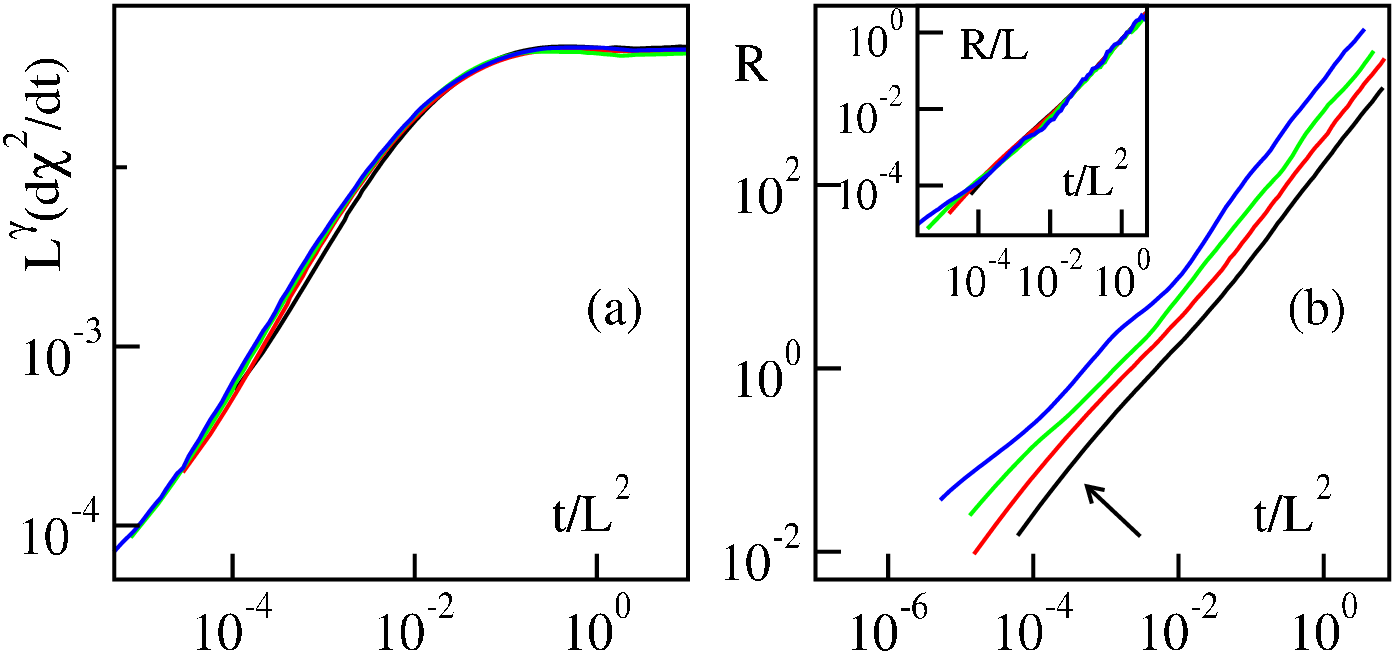}}
\caption{(a) Rescaled FTLE fluctuations for the stochastic model in Eq.~\eqref{stoch},
according to the ansatz \eqref{ansatz_chi}, for
$L=256$, 512, 1024, and 2048 ($\gamma=-1$). (b) The ratio $R$ plotted
for the same model and system sizes. The rescaled ratio $R/L$ is
plotted in the inset.}
\label{fig:fig3}
\end{figure}

The scaling behavior of $\chi^2$ for the model~(\ref{stoch}) is reported in
Fig.~\ref{fig:fig3}(a) (simulations have been performed for an average
{\color{black} $\bar E=\sum_iE_i/L$}
equal to 2), where we see a scenario quite similar to that of the
$\Phi^4$ model with a $\gamma$-value close to $-1$. The close correspondence is
further strengthened by the analysis of $R$ displayed in Fig.~\ref{fig:fig3}(b),
which confirms that the additional rescaling $R\to R/L$ is needed to ensure a
good data collapse.

Unfortunately, a simple recipe to generate a stochastic process with
the {\color{black} correlations}
expected for the FPU model is not available. We
nevertheless believe that there is a compelling evidence that the anomalous
divergence of the diffusion coefficient emerges from the space-time correlation
of the multipliers in tangent space, or, equivalently, of the noise in the
rough-interface picture. It is interesting to notice that the anomaly is stronger
(i.e.~$|\gamma|$ is larger)  in the $\Phi^4$ model which, thermodynamically, is
known to be characterized by a ``normal" (finite) thermal conductivity. The
reason for this seemingly odd conclusion is that the origin of the anomalous
Lyapunov dynamics resides in the hydrodynamic behavior of the multipliers: the
normal diffusion observed in the $\Phi^4$-model is slower than the ``anomalous''
superdiffusion arising in the FPU-$\beta$ context!  

\paragraph{Large-deviation theory.--}
In order to fully appreciate the role of FTLE fluctuations, it is convenient to
introduce the probability $P(\lambda,t,L)$ to observe $\lambda$ over a time $t$
in a system of size $L$. The theory of large deviations predicts
that~\cite{touchette,grassberger88,sepulveda89} (see also \cite{Pikovsky})
\begin{equation}
P(\lambda,t,L) \sim  {\mathrm e}^{-t S(\lambda,L)} \; ,
\end{equation}
where the entropy $S(\lambda,L)$ is a dynamical invariant which has typically a
quadratic minimum at $\langle\lambda\rangle$ (the true LE). The  diffusion
coefficient $D$ is the inverse of the second derivative of $S(\lambda,L)$ in
$\lambda$ at $\lambda=\langle\lambda\rangle$ (see e.g.~\cite{kuptsov11}). Equivalently, one can look at the problem
in terms of the generalized LEs $\mathcal{L}(q)$, defined from the growth rate
of $q$th order moments~\cite{fujisaka84,benzi85}
\begin{equation}
\mathcal{L}(q,L)=q^{-1} \lim_{t\to\infty} t^{-1} \ln \langle \| {\bm{\delta u}}\|^q \rangle \; .
\end{equation}
The two representations are connected via a Legendre transform \cite{Pikovsky}.
Since, to lowest order in $q$~\cite{fujisaka84},
\begin{equation}
\mathcal{L}(q,L) =\langle\lambda\rangle+q D(L)/2 + O(q^2) \; ,
\label{eq:power}
\end{equation}
a divergence of $D$ with $L$ implies the existence of a singularity of
$d\mathcal{L}(q)/dq$ at $q=0$. 
Preliminary computations of higher-order cumulants suggest that 
the divergence is exclusive of the linear term in Eq.~(\ref{eq:power}) due to the
presence of nonanalytic terms in the thermodynamic limit of $\mathcal{L}(q,L)$. 
More refined computational efforts are required to clarify this point.

\paragraph{Conclusions.--}
We have shown that in Hamiltonian models the variance of the maximal
FTLE diverges in the thermodynamic limit. {\color{black} This follows from the 
slow, hydrodynamic fluctuations that affect the local multipliers.
Given the universality classes identified while studying
heat conductivity~\cite{Lepri16}, a similar scenario is expected for the
Lyapunov fluctuations. In particular, different scaling exponents are expected 
in the asymmetric FPU-$\alpha$ model.
Interestingly, the divergence is stronger and qualitatively
different in models exhibiting normal transport such as the
$\Phi^4$ model. In that case, the structure of the Lyapunov vector surface is
not self-affine (i.e.~based on a single Family-Vicsek scaling ansatz): 
two different $\alpha$-exponents must be introduced to describe the growth of the
interface width and sample-to-sample fluctuations, respectively. In consonance, 
the structure factor $\Sigma(k)$ exhibits increasing fluctuations at low $k$. 
Altogether, we expect the divergence of $D$ to carry over to two- and 
three-dimensional setups, where normal diffusion is even more universal. At low 
temperatures the hydrodynamic behavior competes with the intrinsic slowness of 
the dynamics itself: whether this can lead to further anomalies is unknown. 
Also, we may conjecture the existence of dissipative systems subject to certain 
conservation laws that may lead to diverging FTLE fluctuations.}

In the context of roughening processes, the diverging fluctuations reported here
would manifest as a divergence of the fluctuations of the interface velocity. In the
past, various noises with a slow decay of either spatial or temporal
correlations have been studied and different scaling exponents introduced to
characterize the interface evolution~\cite{medina89} (see also \cite{Barabasi}).
However, to our knowledge no systematic investigation of combined
spatio-temporal correlations has been carried out. This type of combined
correlations seems to be essential for the
scenario discussed in this Letter to occur.

Finally, regarding the physical meaning of a diverging diffusion coefficient
$D$, note this does not imply the violation of the central limit theorem, as for
any finite system size $D$ remains finite and thereby the LE well defined.
Nevertheless, intermittent phenomena in tangent-space dynamics become
increasingly important in the thermodynamic limit. How large are the deviations
from a Gaussian approximation is not, however, clear: a complete study of
large deviations is required.

\begin{acknowledgments}
DP acknowledges support by MINECO (Spain) under a Ram\'on y Cajal fellowship.
We acknowledge support by MINECO (Spain) under project No.~FIS2014-59462-P.
\end{acknowledgments}


%

\end{document}